\documentclass[11pt,a4paper]{article}
\usepackage[utf8x]{inputenc}
\usepackage[T1]{fontenc}

\usepackage{helvet}

\usepackage{xcolor} 
\usepackage[pdftex]{graphicx} 
\usepackage[pdftex,linkcolor=black,pdfborder={0 0 0}]{hyperref} 
\usepackage{calc} 
\usepackage{enumitem} 

\usepackage[square,numbers,sort]{natbib}
\usepackage[a4paper, lmargin=0.1\paperwidth, rmargin=0.1\paperwidth, tmargin=0.1111\paperheight, bmargin=0.1111\paperheight]{geometry} 

\usepackage[all]{nowidow} 
\usepackage[protrusion=true,expansion=true]{microtype} 

\usepackage{lipsum} 

\usepackage{titlesec}

\usepackage{color,soul}

\hypersetup{ 	
pdfsubject = {},
pdftitle = {},
pdfauthor = {}
}

\renewcommand{\textcolor}[2]{#2}

\title{ 
\textbf{Reshaping Undergraduate Computer Science Education in the Generative AI Era} (Version 1.0)}
\author{This whitepaper is the report of the NUS-Google Workshops \\ Organized by the AI4SG Lab, NUS School of Computing, SoC Horizons Office, and Google}
\date{}

\begin{document}

\maketitle






\section*{Executive Summary}

Generative AI represents a turning point for Computer Science (CS) education \cite{denny2024computing}. In recent decades, post-secondary CS education has largely focused on what has been seen as practical software engineering skills: implementation-level programming, debugging, testing, and software design, analysis, and documentation. However, this framing is becoming less tenable as generative AI automates many of these tasks, challenging their centrality in CS education.

Recent findings highlight a growing gap between what universities teach and what the software industry needs \cite{clear2025ai, prather2025beyond}. AI coding tools like Claude Code, OpenAI Codex, and Google Gemini can now automate and streamline labor-intensive tasks traditionally assigned to junior roles, such as fixing simple bugs, unit testing, maintaining the build system, writing boilerplate code, and documentation. This shift is disrupting the job market for new graduates, reducing the number of entry-level roles. While the extent to which this contraction is driven by AI versus broader macroeconomic factors (e.g., post-pandemic hiring corrections, rising interest rates) remains an open empirical question, the trend is widely acknowledged across industry and academia. Indeed, the AI-driven transition requires the workforce to develop skills to work and communicate with AI proficiently, effectively, and ethically.

To keep pace with advances in AI technology, CS curricula should consider a shift toward understanding and verifying AI-generated artifacts.
This white paper outlines the findings of two international NUS-Google Workshops (November 2025\footnote{https://futuredu.ai4sg.org/} \& January 2026\footnote{https://futuredu.student.ai4sg.org/}) in Singapore, where we convened faculty members, industry practitioners, and students, and proposes a strategic response to reshape how CS should be taught at the undergraduate level. Based on the findings, we identify critical skills that must be preserved and those that are becoming less important. By incorporating these skills as "breadcrumbs," we can provide helpful nudges and engaging exercises within the current curriculum, enhancing learning experiences for everyone.
We believe that to effectively prepare future computer science graduates, capable of creating, solving problems, and managing, as well as co-creating, artifacts with AI. It is important to consider a shift in curricula. Emphasizing system design, abstraction, and critical evaluation could greatly enhance their education and readiness for the challenges ahead. We propose prerequisites for solutions to reform CS education by fostering AI-native competencies, re-centering fundamental education, enhancing advanced pathways, embracing new pedagogies, and shifting institutional support.

\newpage

\tableofcontents

\section{Background}
\subsection{The Historical Context of Computer Science Pedagogy}
The current disruption in CS education can be understood within the historical context of code scarcity. Traditionally, the curriculum was built on the premise that writing a program was a labor-intensive manual translation of logical intent into rigid syntactic structures \cite{dijkstra1978foolishness, gries1974what}. Consequently, pedagogy emphasized translating ideas into code (or programming mechanics), alongside core foundations such as data structures, algorithms, and system-level abstractions. The implicit assumption was that the struggle to master programming logic was the primary vehicle for developing computational thinking.

Over time, the emphasis has shifted from programming to software engineering. 
Whereas programming is the discrete act of producing code to solve a specific problem, software engineering involves systematic development, operation, and maintenance of software over time \cite{swebok2024}.
This includes maintaining codebases, managing complexity at scale, and navigating changing requirements and human collaboration.



Notably, one strand of CS pedagogy that has always operated above the level of syntax is the teaching of type systems and formal verification. Type systems train students to reason about program correctness through abstraction---specifying what a program should do rather than how it produces output---and have long served as a pedagogical bridge between mathematical reasoning and practical software construction. Similarly, formal verification shows that machine-checkable proofs of correctness can provide guarantees that no amount of testing or code review can match. As AI-generated code proliferates, these techniques become rather \textit{more} important: when humans are no longer the primary authors of code, the ability to specify intended behaviour precisely and to verify that implementations conform to those specifications is arguably the most durable technical skill a CS graduate can possess. Yet type systems and verification have historically been treated as advanced or elective topics rather than foundational ones.

Furthermore, new software systems will include AI components that are non-deterministic and may be trained on data distributions that change over the software's deployment lifetime. Evaluation and maintenance of such software systems are typically not taught in standard undergraduate curricula, even as elective topics. The current disruption is occurring rapidly, making these curricular gaps urgent to address.


\textcolor{blue}{\subsection{Concurrent Initiatives and Positioning}}
\label{concurrent-initiatives}
\textcolor{blue}{This white paper builds on a rapidly growing body of concurrent initiatives addressing the impact of generative AI on CS education. The ACM/IEEE-CS/AAAI \textit{Computer Science Curricula 2023} (CS2023)~\cite{cs2023curriculum}, endorsed in early 2024, represents the current standard for undergraduate CS curriculum guidelines. CS2023 expanded AI as a core knowledge area---from zero Tier-1 hours in CS2013 to substantial required coverage---and introduced a competency-based framework alongside the traditional knowledge model~\cite{eaton2024aics2023}. Our proposals build upon and extend this framework, particularly in articulating a ``verification spectrum'' as a graduate competency that CS2023 does not explicitly foreground.}

\textcolor{blue}{Several parallel efforts overlap with the scope of this white paper. The UC San Diego GenAI in CS Education Consortium\footnote{https://www.teachcswithai.org/}, offers courses integrating GenAI into CS curricula and has been adopted across North America, Europe, and Africa. The ACM GenAI Task Force\footnote{https://acm-education-genai-task-force.github.io/} published its formal report documenting approaches to teaching programming in the GenAI era. The Computing Research Association's LEVEL UP AI Report (September 2025)\footnote{https://cra.org/new-cra-level-up-ai-report-charts-a-path-to-expand-ai-education-nationwide/}, produced through 32 roundtables with over 202 experts, charts strategies for expanding AI curricula in the United States.}

\textcolor{blue}{Our white paper offers several contributions that complement these initiatives. First, our dual-workshop methodology captures both faculty and student perspectives, revealing areas of consensus and divergence that single-stakeholder approaches miss. Second, this paper proposes criteria for program-wide restructuring across all four undergraduate years.}

\subsection{AI Disruptions on Computer Science Education}
The advancement of LLMs and AI-powered coding assistants has altered this landscape \cite{prather2023robots}. Tools such as Claude Code, OpenAI Codex, and Google Gemini provide automated support for many core tasks that junior engineers\footnote{Throughout this white paper, we use the term \textcolor{blue}{\textit{``junior engineer''} in two related senses. In industry, it refers to an entry-level software engineering role whose responsibilities centre on well-scoped, lower-complexity tasks: fixing simple bugs, writing boilerplate code, generating test cases, maintaining build systems, and producing documentation. As a metaphor for AI, it captures the observation that current AI coding tools can perform these same tasks with increasing competence, yet consistently fall short on higher-order skills such as system architecture, verification, security reasoning, and long-term maintainability~\cite{amasanti2025impactaigeneratedsolutionssoftware}. The educational implications of this dual usage are significant: if AI occupies the functional role of a junior engineer, then the traditional apprenticeship pathway through which graduates developed expertise by performing precisely these tasks is disrupted.}} have previously performed \cite{shihab2025effects}.

Some early empirical studies suggest that AI-assisted code generation can yield productivity gains for certain tasks. For instance, Peng et al.~\cite{peng2023impactaideveloperproductivity} found that developers using GitHub Copilot completed an experimental HTTP server task approximately 55\% faster than the control group, though these findings are context-dependent and may not generalise across all development activities.
Routine tasks such as writing boilerplate code, refactoring simple functions, generating test cases, and debugging common errors are now handled by AI with increasing competence.
As AI increasingly automates the translation of logic into programs, the bottleneck of software development is shifting. Software engineers can now offload the burden of rote syntax to AI~\cite{10.1145/3756681.3757081}, allowing them to focus their efforts on higher-order engineering skills.

\textcolor{blue}{However, productivity gains from AI tools may be context-dependent and subject to systematic overestimation. A randomized controlled trial by Becker et al.~\cite{becker2025metr} found that experienced open-source developers actually took 19\% longer to complete real tasks when using AI tools, despite believing they were 20\% faster, which shows a striking gap between perception and reality. These findings suggest that the productivity benefits of AI coding tools may be most pronounced for less experienced developers or less complex codebases, rather than uniformly applicable.}
Emerging empirical evidence suggests that the productivity gains from AI coding tools may come at a significant cost to software quality and maintainability. He et al. \cite{he2026cursor} conducted a large-scale causal study of Cursor adoption across open-source GitHub projects using a difference-in-differences design. While they found that Cursor adoption led to a statistically significant increase in short-term development velocity, this speedup was transient. More critically, adoption was associated with a substantial and persistent increase in static analysis warnings and code complexity: factors that the authors identify as drivers of long-term velocity slowdown. These findings underscore a central tension: AI tools can accelerate code production, but without rigorous human oversight, the resulting code accumulates technical debt that erodes the initial productivity gains. 


Thus, the central challenge for CS education is not whether students will use AI, but how they will learn to work with and supervise it effectively.    
Doing so still requires today's students to learn a strong conceptual understanding of computation and software systems. 
Historically, code production served as a mechanism for students to develop, exercise, fine-tune, and demonstrate their understanding of and ability to apply these concepts. 
Today, however, as AI systems increasingly generate code, students get fewer and fewer of these experiences. 

Most critically, students' use of AI is undermining their intellectual skills.
When a class assigns a programming assignment to a student who uses AI to generate the program (whether instructed to do so or not), the student gets less practice in problem-solving and critical thinking, and research shows that these habits disappear from students' practices remarkably quickly~\cite{choudhuri2026cantthink, fan2025metacoglazy, choudhuri2025genAIstudents, tankelevitch2024metacognitive, lee2025impact}.
If universities graduate students without the skills to think critically and problem-solve about software systems (whether written by AI or not), how can they responsibly manage such systems?
That is the problem that universities must now address.
\textcolor{blue}{This concern is supported by recent causal evidence. Bastani et al.~\cite{bastani2025guardrails} conducted a randomized controlled trial with nearly 1,000 students and found that unrestricted GPT-4 access, while improving practice performance by 48\%, led to a 17\% decline in exam scores when AI was subsequently removed. Crucially, a guardrail version designed to provide hints rather than direct answers mitigated these negative effects---suggesting that the pedagogical design of AI integration, not AI use per se, determines learning outcomes.}


\newpage
\section{Methodology: The Workshops}
In response to advances in AI, we sought to gather voices from a variety of stakeholders in CS education. In the first workshop, we brought together CS faculty members and IT industry practitioners from around the world to identify problems, assumptions, and opportunities arising from the advent of AI, and to brainstorm the fundamental needs and solutions for the future of CS education. The second workshop included current CS students and recent graduates in CS-related fields as participants to solicit views from junior practitioners and to understand their lived experiences as the generation most impacted by this AI paradigm shift.

\begin{figure}
    \centering
    \includegraphics[width=10cm]{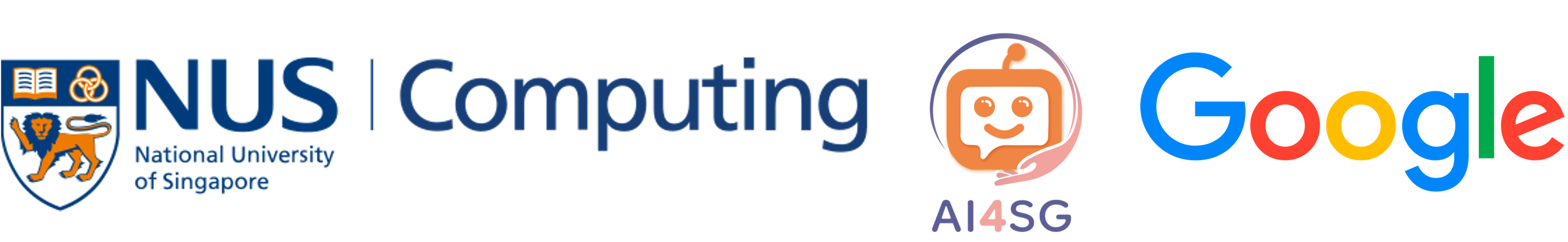}
    \caption{Our workshops are initiatives by the \href{https://www.ai4sg.org/}{AI4SG Lab}, at the NUS School of Computing, and Google}
    \label{fig:logos}
\end{figure}

\subsection{First Workshop: Faculty-focused Version}

In November 2025\footnote{https://futuredu.ai4sg.org/}, we first organized a three-day workshop that brought together CS faculty members and IT industry practitioners from around the world. The workshop had 30 attendees, with a diverse range of participants from North America, Asia, Europe, and Oceania.

\textbf{Day 1:} The workshop commenced with a warm-up discussion for early arrivals on the first day. This session was dedicated to identifying foundational concepts that have become increasingly critical, as well as subjects that may now be considered ``obsolete'' or warrant reduced emphasis. Specifically, attendees were divided into three separate groups to discuss the topic across four categories: KEEP, CUT, GAP, and THREAT.

\textbf{Day 2:} Moving into the main workshop on the second day, the objective shifted to architecting the core content ("what") of the new curriculum, progressing from problem statements to defining guiding principles and developing concrete models. The attendees were divided into five groups to address a general question: \textbf{What are the core competencies for a Computer Science graduate in 2030?} Each group were also assigned theme-specific questions. Group members were assigned based on expertise and diversity to ensure a range of perspectives were represented in each group.

\begin{itemize}
    \item Foundational \& Theoretical Skills (Group 1): \textit{What core CS theory, topics, or knowledge cannot be replaced?}

    \item The New Software Engineer Role (Group 2): \textit{What practical skills (code review, systems integration, testing, prompt engineering) define the new SE role?}

    \item AI-Native Competencies (Group 3): \textit{What new, specific AI skills are required (e.g., applied ML, AI systems design)?}

    \item Professional \& Ethical Skills (Group 4): \textit{What soft skills (ethics, security, communication, collaboration) are now paramount?}

    \item New Pedagogies \& Teaching Strategies (Group 5): \textit{How and what we teach to foster these new competencies?}
\end{itemize}

\begin{figure}
    \centering
    \includegraphics[width=0.8\linewidth]{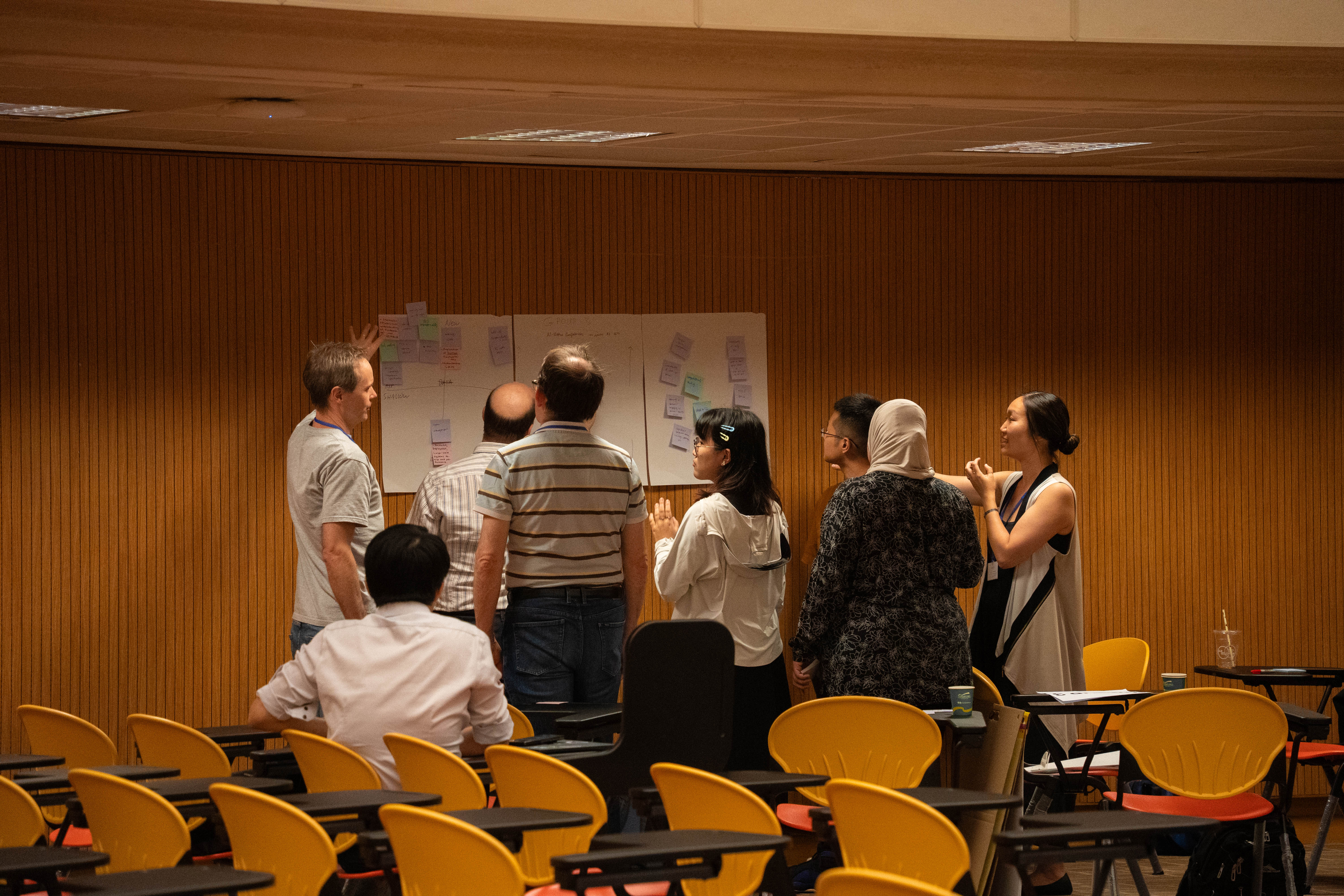}
    \caption{Photo from Day 2. The faculty workshop features multiple rounds of group brainstorming activities.}
    \label{fig:faculty-ws-snapshot}
\end{figure}

\textbf{Day 3:} The final day of the workshop continued the discussion of topics stemming from the conclusions reached on Day 2. The results revealed overlapping outcomes among groups. Consequently, we reorganized the groups to facilitate a more focused discussion. The central objective of the discussion was to determine the structure of the four-year curriculum, drawing upon the established ``Graduate Profile'' as the guiding framework.

\begin{itemize}
    \item Group 1: The New Core (Years 1-2): This group redesigned the introductory sequence. The group brainstormed solutions for teaching CS101 in the era when AI can write the code.

    \item Group 2: Advanced Pathways (Years 3-4): The group was assigned to propose solutions for teaching computer science for new advanced tracks or specializations.

    \item Group 3: The AI-as-Collaborator Policy: The group was assigned to define AI's role in education, the AI usage policy, and new assessment methods that are AI-resistant or AI-inclusive. 

    \item Group 4: The Ethics \& Societal Impact Thread: The group designed a guideline for ethics, security, and societal impact that integrates into every year of the curriculum, beyond just one course.

\end{itemize}

In the afternoon session on Day 3, attendees gathered in groups based on their topics of interest to draft proposals to reform the CS curriculum.

\begin{itemize}
    \item Foundational Research for the Next-Generation SE (Group 1): Aligned with Day 1's "Foundational Skills" and "New SE Role", this group defined the new science required to build, verify, and secure the complex AI-native systems that graduates will be asked to create (e.g., verifiable systems, AI-driven SE, ethical auditing).

    \item The Co-evolving Talent \& Infrastructure Pipeline (Group 2): This group redesigned the academia-industry interface beyond internships to create a dynamic, shared ecosystem for both talent and technology (e.g., new collaborative models, shared infrastructure, continuous feedback loop).

    \item The Research Agenda for AI-augmented Pedagogy (Group 3): This group moved beyond a simple "best practices" guide to define the fundamental research needed to build, deploy, and understand AI's role in education (e.g., cognitive impact, assessment innovation, co-educator development).

    \item The Faculty \& Institutional Action Plan (Group 4): To complement Day 1's vision, this group envisioned institutional change and faculty development, together with potential barriers, policies, incentives, and resource allocation.
\end{itemize}

\subsection{Second Workshop: Student-focused Version}

As a follow-up, in January 2026\footnote{https://futuredu.student.ai4sg.org/}, we convened a one-day workshop to explore the evolving landscape of computer science education in the age of AI. The event brought together 60 participants, including students and recent graduates from across the globe, with approximately half of the cohort representing Singapore-based institutions. This diverse mix of local and international perspectives provided a comprehensive foundation for tackling the complexities of modernizing CS curricula. Attendees were divided into eight working groups (Groups 1–8) assigned to four critical discussion themes:
\begin{itemize}
    \item Challenges \& Response (Groups 1-2): Focused on disruption and institutional adaptation.

    \item CS Education Fundamentals (Groups 3-4): Targeted the foundational undergraduate years 1 and 2.

    \item Advanced Pathways (Groups 5-6): Addressed the specialized needs of undergraduate years 3 and 4.

    \item Profile of the Modern CS Graduate (Groups 7-8): Concentrated on graduate-level competencies and industry alignment.
\end{itemize}

To ground the sessions, each group was provided with a tailored set of questions based on the findings from our inaugural workshop. These prompts guided the attendees through intense brainstorming and collaborative problem-solving, culminating in a series of group presentations where each team shared their synthesized findings.

\textcolor{blue}{\subsection{Limitations}}

\textcolor{blue}{We view these workshops as a starting point for an ongoing, evidence-driven conversation rather than a definitive conclusion. However, we acknowledge several limitations of our workshops. First, workshop participants were self-selected and likely already engaged with questions of AI in education, introducing potential selection bias. Second, while the faculty workshop included international participants from North America, Asia, Europe, and Oceania, the student workshop drew approximately half its cohort from Singapore-based institutions, limiting geographic generalizability. Third, workshop findings reflect participants' perceptions and proposals rather than empirically validated outcomes; the proposed solutions remain to be tested. Fourth, industry perspectives were represented primarily through individual practitioners rather than systematic employer surveys, and a broader range of industry input would strengthen the findings.}

\newpage

\section{Summary of Outcomes}

\subsection{Faculty Workshop}


\subsubsection{Capabilities of Future Software Engineers}

The discussion on Day 1 focused on preparing students for a world where AI has the capabilities of a "junior engineer", as defined in Section~\ref{concurrent-initiatives}. Participants generally agreed that the educational focus should shift towards higher-level skills. Three key examples raised were abstraction, verification, and taking responsibility for systems. 

\begin{itemize}

\item \textit{Abstraction}: Abstract computational thinking gains importance when code production becomes cheaper. The ability to understand computational structures that span programming languages, architect systems, and analyze complex AI models and their limitations is what differentiates skilled software engineers from mere programmers.


\item \textit{Verification:} Participants agreed that AI models effectively act as junior software engineers, capable of producing code but often making suboptimal architectural decisions. Verification thus serves as a crucial skill in which humans play the role of a "senior software engineer". Importantly, ``verification'' in this context encompasses a spectrum of rigor: from informal code review and testing, through static analysis and type-system-guided development, to full formal verification using proof assistants. \textcolor{blue}{This verification spectrum aligns with the three-level framework proposed in the formal methods education literature~\cite{fmthinking2024}, which argues that such thinking can be embedded within existing courses without requiring additional curricular expansion.}

\item \textit{Taking responsibility for systems:} With AI lowering the barrier to building systems, participants raised the legal, moral, and ethical responsibilities associated with such systems as issues that remain squarely in the human domain. Ethical responsibilities include designing systems for human betterment, such as AI augmentation over replacement. Legal responsibilities include potential accreditation and liability for systems, particularly when AI generates code for safety-critical applications. As systems become increasingly easy to produce through automation, the need for deliberate, thoughtful, and responsible system design and production gains renewed importance. 

\end{itemize}

The consensus was that as creating code becomes cheaper, the ability to discern quality, ensure security, and maintain ethical standards becomes even more valuable. Teaching students to operate across the verification spectrum is also essential if graduates are to take genuine responsibility for AI-generated systems. Consequently, the traditional pathway in which junior engineers gradually build expertise through progressively complex tasks may need to be compressed or restructured. Students will likely need to engage with architectural thinking and integration challenges earlier in their education, though doing so effectively will require carefully scaffolded learning experiences rather than a simple bypass of foundational stages. 

\subsubsection{Reshaping the Curriculum}


\textbf{Shifting the Core (Years 1-2).}
The key shift is to move introductory CS learning goals from low-level syntax and rote coding to higher-order computational thinking grounded in computing fundamentals, while focusing human-centric skills that cannot be automated. Key features include, but are not limited to:

\begin{itemize}

\item \textit{Embracing AI integration:} AI tools are intentionally integrated from the start to enable students to work on larger, more motivating projects, thereby reducing low-level workload and avoiding early student demotivation.

\item \textit{Computational Thinking and theory:} The new core seeks to foster fundamental and theoretical skills, emphasizing Computational Thinking as its foundation. It focuses on the enduring importance of core CS theory, such as abstraction, decomposition, and algorithmic reasoning, and fundamentals such as problem-solving, scientific reasoning, and learning agility. 
It proposes teaching up the abstraction ladder, allowing students to delegate low-level coding tasks to AI (after learning the fundamentals) and instead focus on problem specification and problem solving. Type-driven development and lightweight formal specification (such as preconditions, postconditions, and contracts) offer an analogous pedagogical mechanism for achieving this shift: they train students to express intent precisely and to reason about correctness at the specification level rather than the syntax level, which is a similar type of skill required to direct and verify AI-generated code effectively. 

\item \textit{Human-centric skills:} At the same time, crucially, soft skills such as communication, collaboration, ethics, and societal reasoning are important core CS competencies, rather than peripheral courses. These are vital for understanding and communicating ``what to build'' and its broader implications; students must be equipped to justify systems and assess their impact, rather than just implementing them.

\end{itemize}


\textbf{Integrating AI into Advanced Pathways (Years 3-4).}
Participants agreed that AI could be a thread woven throughout the curriculum, not a standalone add-on. This vision focuses on upgrading existing specializations, such as human-centered computing, security, and computer architecture, by adding AI-focused components, resulting in "AI-enabled" or even "AI-native" versions of these tracks. Detailed examples included:

\begin{itemize}

\item \textit{The new software engineering role in AI:} This addresses the practical skills needed for the evolving software engineering profession. It advocates for AI-ready assignments and experiential learning, proposing that students work on large, real-world, open-source projects that utilize AI tools. Such assignments could also incorporate verification-oriented workflows, where students use AI to generate code and then apply static analysis, property-based testing, or formal verification tools to validate the output—mirroring the quality assurance challenges that He et al.\ \cite{he2026cursor} have shown to be a major bottleneck in real-world AI-assisted development. Students will also need to learn how to incorporate AI components into software systems and evaluate their effectiveness \cite{Reddi2026MLSystems}.

\item \textit{Mathematical foundations and Responsible AI:} Strengthening connections between mathematics and AI ensures students understand the technical basis of responsible AI. Topics such as probability, linear algebra, and discrete mathematics, often taught as core courses, could be explicitly linked to AI applications rather than taught in isolation. At the same time, educators can emphasize ethical awareness and frame mathematical training to support responsible AI practices while recognizing that technical solutions alone cannot address all ethical challenges.

\item \textit{Evaluation of AI tool usage:} Attendees highlighted the importance of assessing the process of how students interact with AI for coding and the problem-solving process, shifting away from merely evaluating the final product. The end goal is to produce employable graduates with skills that differentiate them in the AI-supported workforce.

\end{itemize}

\begin{figure}
    \centering
    \includegraphics[width=\linewidth]{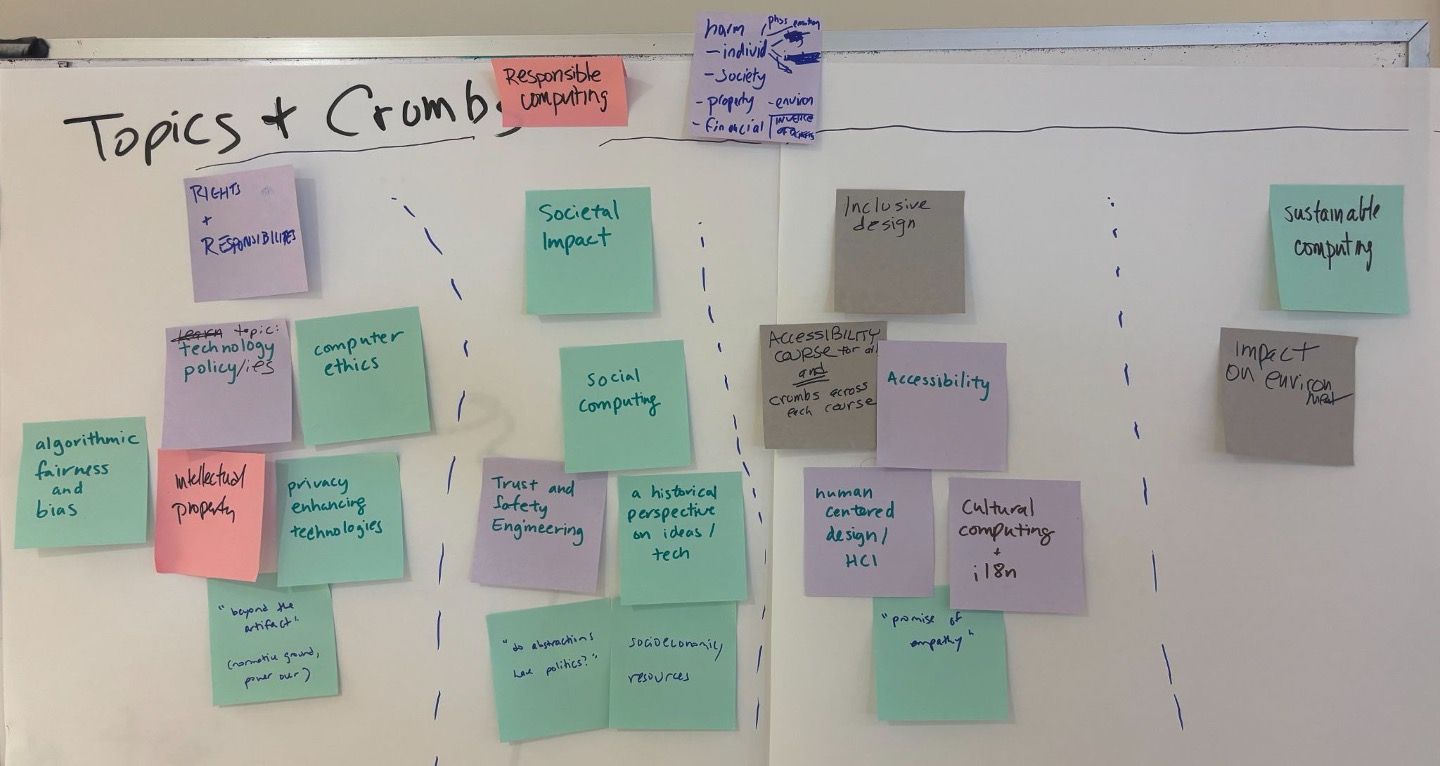}
    \caption{Faculty participants proposed the Breadcrumb strategy as a lightweight solution to foster students' AI-native skills. Small breadcrumbs (i.e., skills or topics) can be integrated progressively throughout the curriculum using small practical exercises, prompts, or nudges. This figure shows a part of the AI-native skill set brainstormed in the workshop.}
    \label{fig:breadcrumb}
\end{figure}

\textcolor{pink}{\textbf{Embedding Skills through the Breadcrumb Strategy.} Skills such as ethics, social \& ecological impact, and responsible AI can be integrated throughout the four-year curriculum. Workshop attendees proposed the ``Breadcrumb'' strategy, which integrates small, practical exercises, nudges, or prompts into multiple courses --- through assignments, exams, or projects. It is aimed at building students’ ability to evaluate impacts, follow accessibility and legal standards, and make responsible design choices. In practical implementation, Breadcrumbs can be sequenced to build on previous knowledge, reinforcing learning year by year, and designed to fit diverse courses (theory to software engineering).} 


\subsubsection{Challenges with Implementing Curricula Change}



\textbf{Supporting Faculty and Institutional Action Plans.} The key to encouraging adoption is faculty engagement. Strategies such as opt-in participation and incentives (e.g., recognition, small funding) may help mitigate resistance. It is crucial to address instructor skepticism by framing AI adoption as low-risk experimentation. Effective adoption requires significant institutional support and resourcing. This includes securing budget for AI tools, providing dedicated staff and lab technicians to assist faculty, and ensuring management buy-in by aligning AI integration with core institutional priorities, such as student outcomes and reputation.


To maintain relevance, curriculum adaptation and teaching strategies must evolve. Faculty should be encouraged to rethink course content as AI automates certain skills, to engage in practical experimentation with AI-assisted assignments, and to utilize resource repositories to share best practices. In addition, industry collaboration and resource sharing were deemed vital, advocating for a consortium-based approach to share computational resources and emerging technologies. Encouraging companies to support open access for education can help universities bridge the gap between resource constraints and the fast pace of AI development.
Finally, the impact on student experience and institutional outcomes requires careful management. This involves striking a balance between teaching innovation and evaluating metrics, while ensuring that AI integration maintains high educational quality and prepares students for a rapidly transforming job market.

\subsection{Student Workshop}

\textbf{The Evolving Role of AI and Student Concerns.} 
Student participants viewed AI as a multifaceted tool, functioning as an accessible 24/7 co-pilot, partner, and mentor for tasks such as academic writing, prototyping, simplifying complex topics, and accelerating workflows. It enables personalized and on-demand learning. However, the groups expressed significant concern about the erosion of critical thinking, overreliance, and student laziness. Concerns also centered on academic dishonesty (plagiarism, using AI to produce ``fake answers'') and the need to teach students how to verify and reflect on potentially unreliable AI outputs (e.g., hallucinations). The consensus was that AI's role is \textit{fluid}: a shortcut for tasks with clear, predictable outcomes (e.g., information search, simple automation) and a collaborator for uncertain, creative tasks where students must inject their own critical thinking and opinions.


\textbf{Re-prioritizing Foundational and Soft Skills.} The discussions highlighted that CS fundamentals—including strong basic math, data structures and algorithms, debugging concepts, and knowledge of operating systems and networking—have become more essential than ever as a core base for understanding and applying AI-generated code. Alongside these technical foundations, student groups stressed the importance of new cognitive and soft skills. Cognitive skills include learning how to learn, how to ask, and, in a provocative turn, ``how to unlearn.'' Essential soft skills, such as responsibility, ethical thinking, business awareness, humility, and caution, were identified as qualities that truly make a CS graduate stand out, even as AI can handle average-level and advanced technical problems. Students identified a combination of emerging and established technical competencies that are crucial for modern IT roles. Among the new skills highlighted is prompt engineering, although participants noted that its long-term relevance may be uncertain as AI systems continue to evolve. Just as important are well-established software engineering practices that gain renewed significance in the AI era. These practices include specification writing (defining the desired system behavior that will be later transferred into code), critical code review (assessing AI-generated output for correctness, security, and maintainability), and troubleshooting edge cases (distinguishing failure modes that AI tools often overlook). This distinction is important: while these latter skills are not new, their importance in a graduate's profile has increased significantly, especially when code is generated by AI rather than written from scratch.


\begin{figure}
    \centering
    \includegraphics[width=0.8\linewidth]{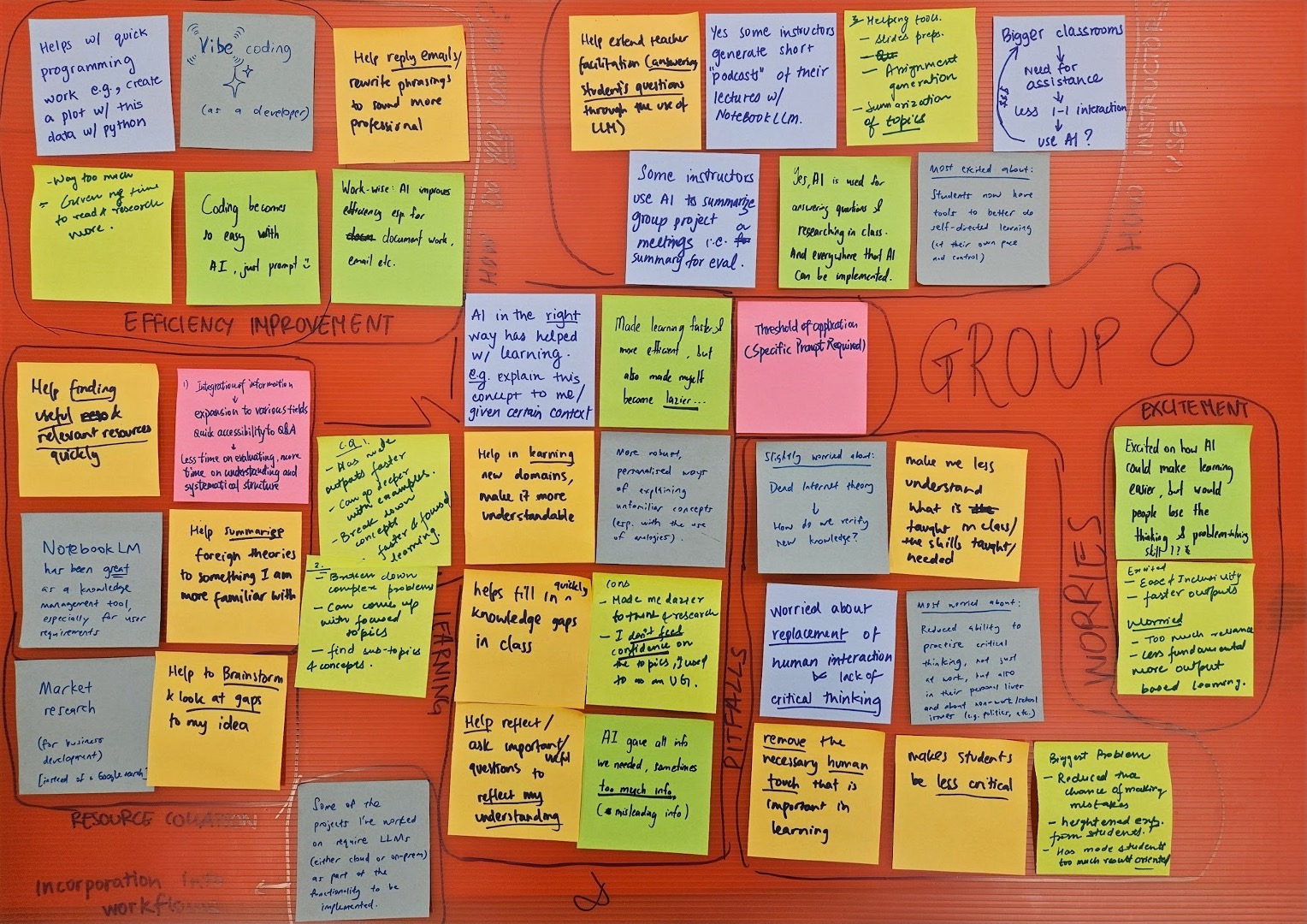}
    \caption{Example brainstorming work from the student workshop. Students voiced opportunities, excitement, and concerns about the advancement of AI with regard to CS education.}
    \label{fig:student-workshop}
\end{figure}

\textbf{Changes to Curriculum and Assessment.} Participants advocated for major shifts in CS education away from traditional models focused on memorizing syntax. Recommendations for a curriculum shift include moving towards project-based learning to apply concepts in real-world, "in the wild" environments, introducing algorithms earlier in the curriculum, and incorporating more explicit instruction on AI ethics, legal compliance, and the social science aspects of technology. For assessment reform, the focus can shift from evaluating only the final answer to assessing the problem-solving process, understanding of errors, and debugging. Proposed assignment formats include presentations, "paper-like" work where students take authorship, and, notably, submissions of the history of their interactions with generative AI to ensure genuine understanding.


\newpage

\section{Consensus and Divergences between Faculty and Students}
Both workshops demonstrated a shared understanding that AI is fundamentally changing the landscape of CS education, necessitating a comprehensive shift in pedagogy and curriculum focus. In this section, we discuss areas of consensus between the two workshops and the nuanced differences between them.

\subsection{Areas of Consensus}
\begin{itemize}
    \item \textbf{Shifting from low-level to high-level.} There was a clear consensus that instruction must move away from memorizing syntax and mechanical coding and pivot toward mastering high-level concepts, abstraction, and problem decomposition.

    \item \textbf{Theoretical foundations to verify AI outputs.}  Both workshops affirmed the enduring importance of core CS theory, such as mathematics, data structures, and algorithms. These foundational skills are deemed essential for a human to critically verify and debug the output generated by AI. Notably, neither workshop drew a clear distinction between informal verification (e.g., code review, testing) and formal or semi-formal approaches (e.g., type checking, static analysis, specification-based reasoning), suggesting that the precise meaning of ``verification'' as a graduate competency remains underspecified and warrants further clarification in curriculum design. 
    
    \item \textbf{Fostering meta skills.} All agreed that new "meta-skills" are now paramount, though they emphasized different aspects: faculty focused on professional, human-centered competencies (i.e., human-computer interaction, communication, ethics, professional liability), while students highlighted personal intellectual skills (for example, critical thinking, self-regulation, and metacognition). 

    \item \textbf{AI-native assessment.} Both workshops recognised a strong shared conviction that assessment practices must evolve to measure authentic understanding in an AI-enabled learning environment, with both proposing a dual approach of controlled (AI-free) testing and open (AI-allowed) assignments.

    \item \textbf{Project-based and experiential learning.} Both workshops converged on the value of engaging students with real-life, larger-scale projects rather than small, isolated programming exercises. Faculty participants proposed ``reverse-engineering'' methods, in which students extract specifications and requirements from existing compiled solutions. This approach serves as a pedagogical process for sense-making, conceptual learning, and verification learning, grounded in real-world codebases. Students advocated for project-based learning applied ``in the wild.'' Though the pedagogical framing differed, the shared conviction that authentic, large-scale projects should anchor the curriculum was a clear area of consensus.

\end{itemize}

\subsection{Differing Areas of Concern}

While unified on the broad direction, the workshops differed in their primary concerns and proposed solutions.

\begin{itemize}
    \item \textbf{Junior-Senior competency gap.} Faculty primarily focused on professional competency gaps between entry-level and senior roles. They mentioned the ``Junior Engineer problem,'' which requires new graduates to immediately possess ``Senior Engineer'' architectural skills. Conversely, the student workshop approached this problem more from an employability standpoint, seeking ways to stand out in an age of AI coding agents.

    \item \textbf{Professional liability of AI.} The faculty workshop voiced institutional challenges like the lack of professional liability for AI-generated code or challenges with assessment and teaching, when AI can perform assessment. The professional liability concern is particularly acute in safety-critical domains, where AI-generated code is introduced into systems without the formal sign-off processes that exist in disciplines such as civil or electrical engineering. 


    \item \textbf{Pedagogical mechanisms.} While both workshops valued project-based and experiential learning, their views were different in how to structure it. The faculty workshop proposed specific instructor-driven curricular formats such as ``prompt-first'' courses and structured ``reverse-engineering'' modules with defined learning scaffolds. The student workshop, by contrast, favored greater student autonomy in selecting real-world problems and learning pathways, with less emphasis on prescribed pedagogical structures. The divergence is thus not about \textit{whether} to use real-world projects, but about the degree of structure and instructor direction in how those projects are framed.
    
\end{itemize}


\newpage

\section{Discussion}

\subsection{What Are AI Disruptions in CS Education?} 

The transformative influence of AI presents challenges to established norms in CS professional progression, the integrity of assessment, and the foundational understanding of professional competence. Faculty participants agreed that the advancement of AI fundamentally disrupts the traditional learning path of computer science students. \textcolor{blue}{As defined in the prior section, ``junior engineer'' in this white paper refers both to the entry-level industry role centered on well-scoped, lower-complexity tasks and to the observation that AI tools now perform these same tasks with increasing competence.} AI systems increasingly function as entry-level ``junior engineers'': they can generate working code quickly, yet they routinely make poor architectural, security, or performance decisions. \textcolor{blue}{This disrupts the historical apprenticeship model in which students gradually build expertise through progressively complex tasks and evolve toward senior responsibility. 

Large-scale evidence supports this concern: Brynjolfsson et al.~\cite{brynjolfsson2025canaries} analyzed ADP payroll data covering millions of U.S.\ workers and found that employment for software developers aged 22--25 declined approximately 16\% relative to less AI-exposed occupations since late 2022. The authors suggest that while adjustments following the pandemic may play a role in earlier declines, the significant impact on entry-level positions exposed to AI persists even when accounting for company-specific factors. This indicates that the distinction between "junior" and "senior" responsibilities is indeed evolving.}
In this context, the importance of low-level syntax drills, memorization of libraries, or even the choice of programming language diminishes, while architectural reasoning, systems thinking, and critical evaluation of AI outputs become central learning outcomes. Evidence is already emerging that this shift is not merely theoretical: He et al.~\cite{he2026cursor} found that Cursor adoption in open-source projects led to a persistent increase in static analysis warnings and code complexity, suggesting that without deliberate quality assurance practices, AI-assisted development accumulates technical debt faster than it generates value.

Even before the rise of AI-assisted development, many computer science students already perceived a gap between what they learned in university and the skills that industry demanded~\cite{clear2025ai}. With the rapid emergence of powerful AI coding tools, this conceptual gap has widened further. Students increasingly observe developers using AI to generate code, debug programs, and explore design alternatives, which can make traditional coursework seem even more removed from real-world practice. At the same time, some companies have become more cautious about hiring junior engineers and, in certain cases, have reduced or even eliminated entry-level positions~\cite{prather2025beyond}, relying more heavily on experienced developers augmented by AI tools. \textcolor{blue}{However, the metaphor of AI as an entry-level engineer warrants careful qualification. Evidence suggests that even experienced developers may systematically overestimate AI's benefits: Becker et al.~\cite{becker2025metr} found a striking gap between perceived and measured productivity in a randomized controlled trial. This perception gap itself becomes an educational concern, reinforcing the importance of teaching students to critically evaluate not only AI outputs but also their own AI-mediated workflows.} This has further reinforced the perception that the entry-level pipeline is shrinking. When students see fewer opportunities for junior engineers and feel that the skills emphasized in school may not match the evolving demands of industry, their motivation to invest deeply in traditional coursework can decline significantly.


Moreover, participants expressed concerns about assessment practices, learning equity, and student self-regulation. Traditional take-home assignments no longer reliably demonstrate a student’s own work, as AI assistance blurs the boundary between independent effort and tool-supported production~\cite{mohammadkarimi2023teachers}. At the same time, invigilated examinations increasingly reveal a widening gap: stronger students tend to use AI as a learning accelerator, whereas weaker students may over-rely on it, developing an illusion of competence that fails under time-pressured conditions. This dynamic highlights a serious deficit in metacognitive skills, with many students unable to accurately assess their own understanding and limitations. Although oral examinations could be a promising alternative for assessing authentic understanding~\cite{Theobold04052021}, the faculty workshop suggested that such approaches are rarely scalable given existing constraints on faculty time and resources. \textcolor{blue}{However, recent innovations in AI-assisted oral examination administration, including systems that generate personalized viva questions and conduct AI-mediated oral defenses at low per-student cost, suggest that this scalability barrier may be diminishing~\cite{ipeirotis2026scalablepersonalizedoralassessments}.} These assessment tensions are further exacerbated by issues of trust, as inconsistent faculty practices around AI use risk undermining the educational relationship between students and instructors.


Beyond coursework, participants identified challenges that extend beyond undergraduate education into professional work and research. As AI-generated code increasingly enters safety-critical systems, the absence of clear professional accreditation and liability frameworks in computer science arguably 
becomes more salient. Unlike fields such as civil engineering, there is no widely accepted model in which a human professional is explicitly accountable for their work, let alone for AI-assisted outputs. Formal verification offers one pathway forward: if a system's correctness can be established through machine-checkable proofs rather than human judgment alone, the question of professional sign-off shifts from ``did someone review this code?'' to ``does this code meet its specification?'' --- a far more tractable standard for AI-generated artifacts. 
These trends point to an urgent need for curricula that emphasize verification, security, performance evaluation, and ethical responsibility, as well as communication and interpersonal skills that remain \textit{uniquely human} and essential to responsible system design.



\subsection{How Do Students Learn CS in the Era of AI?}

The student workshop highlighted a significant shift in how students learn CS in the era of AI. The consensus is that AI is a powerful tool, but its integration requires a re-evaluation of educational priorities to safeguard and enhance human-centric skills.

Student participants reported incorporating AI into their academic journey. For undergraduate students, AI primarily serves as a 24/7 partner for assignment help.
AI is valued for its ability to explain complex topics, provide alternative perspectives, and generate ideas for cross-referencing, effectively lowering the barrier to entry for coding. However, this ease of use sparked significant concern about the erosion of critical thinking, original input, and students' intrinsic motivation, as well as the risk of malicious use, such as plagiarism. \textcolor{blue}{These observations align with experimental evidence showing that the design of AI tools significantly mediates their educational impact: guardrailed AI tutors that provide scaffolded hints rather than direct answers can preserve learning outcomes, whereas unrestricted access creates a ``crutch'' effect that undermines skill acquisition~\cite{bastani2025guardrails}.}

The role of AI is seen as dynamic, changing based on the specific task. The groups distinguished between two main roles. AI could be rather treated as a shortcut for tasks with a clearly defined or desired outcome, such as routine information searches, simple automation, or data collection and consolidation. Conversely, students can treat AI as a collaborator for tasks with uncertain or creative outcomes, such as design thinking or creative writing, where students are required to actively inject their own unique thoughts, opinions, and agency into the final product. Ultimately, the groups concluded that the educational system must adapt assessment methods to account for AI's capabilities, for example, by requiring the submission of AI-use history alongside assignments or by placing greater emphasis on presentations and public peer review to ensure individual responsibility for the work.

\subsection{Nuanced Implementation of Pedagogical Changes}

The differing perspectives of students and faculty highlight the need to involve a range of stakeholders in such discussions. Changes to curricula must address students' emotional and personal concerns, rather than simply meeting institutional objectives such as accurate assessment or graduates' employability. Given that news about "PhD-level AI" \cite{phd2025bbc} and fully-agentic workflows \cite{computer2026perplexity} continues to dominate headlines, current and prospective CS students rightly worry about their relevance and employability in the job market. Acknowledging these concerns is central to ensuring student buy-in and, therefore, active participation in any proposed curricular changes. Framing verification and specification skills not as an additional burden but as tools that give graduates a durable competitive advantage over AI—since machines generate code but cannot yet prove its correctness—may help address students' anxieties about relevance in an AI-dominated job market.

\newpage

\section{Criteria that Any Solution Must Meet}
\label{criteria}
Participants voiced a diverse set of solutions to address the AI disruption in CS education. During the workshop discussion, they noted the requirements that these solutions must meet. We categorized them into four themes: fostering AI-native competencies; shaping strong fundamentals (Undergraduate years 1-2); enhancing advanced pathways (Undergraduate years 3-4); and embracing new pedagogical strategies and shifting institutional support.

\begin{figure}
    \centering
    \includegraphics[width=\linewidth]{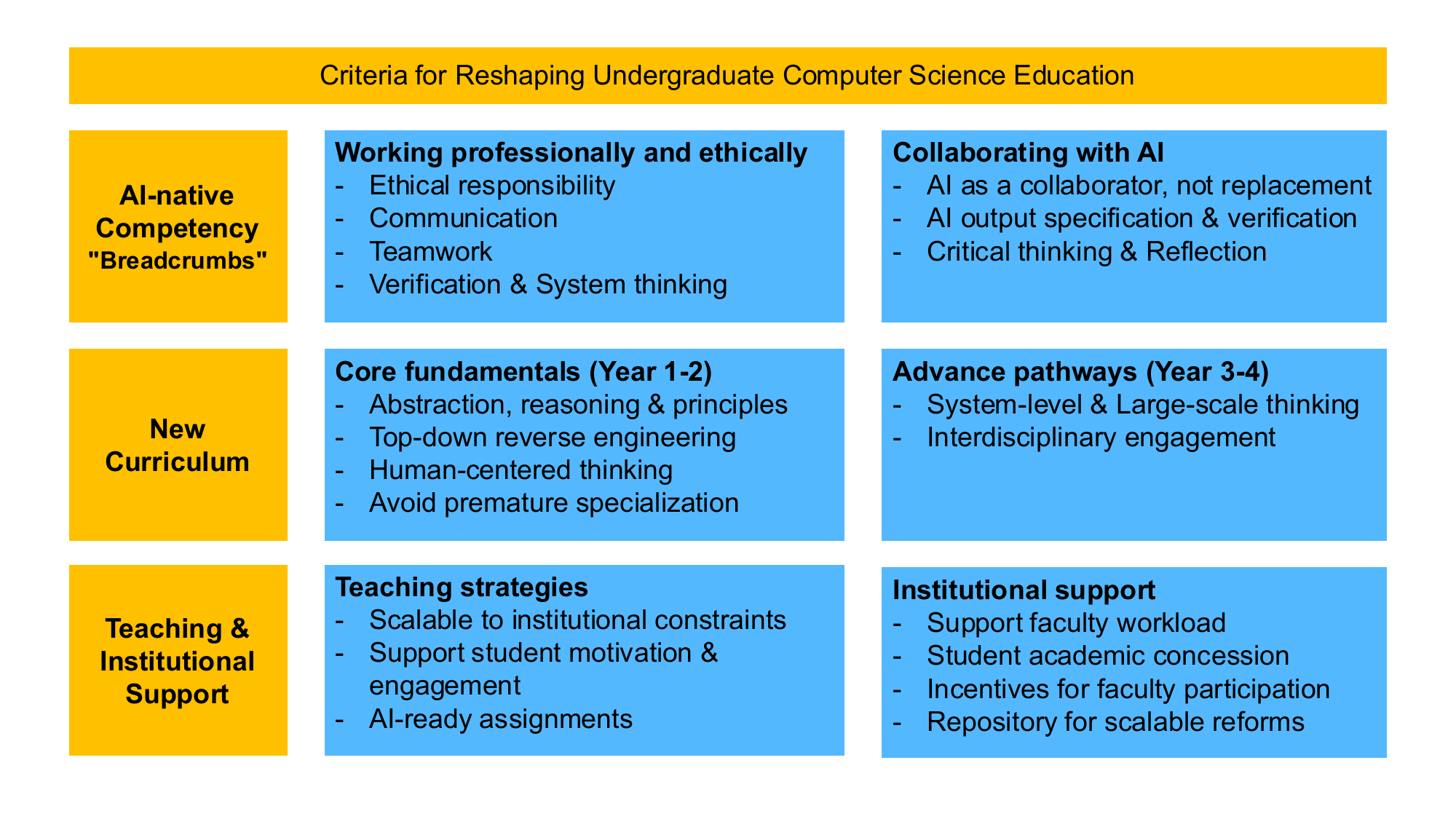}
    \caption{Criteria for CS Education Reforming Solutions}
    \label{fig:criteria}
\end{figure}

\subsection{Fostering AI-Native Competencies}
Workshop participants reached consensus that CS education must train students to work with AI proficiently, responsibly, and ethically. While AI can perform low-level tasks (such as programming) quickly, it may not excel at high-level tasks, which encompass metacognitive thinking, abstraction, and critical thinking. Current AI systems also cannot verify their own outputs against formal specifications or provide guarantees of correctness, making human competence in specification and verification an AI-native skill in itself—not merely a legacy practice.

Particularly, the issue of ethics was raised in the discussion. In the CS context, ethical considerations go beyond abstract discussions of values. It also includes concrete technical aspects, such as recognizing and mitigating bias in training data, evaluating privacy risks in data collection and logging, assessing security vulnerabilities before deployment, understanding when system failures could cause real-world harm, and deciding whether an AI-generated solution is sufficiently reliable for use in safety-critical or high-stakes settings. To ensure that downstream applications do not trigger and amplify societal issues, \textbf{it is important that CS graduates can decide beforehand whether using AI is the right approach and verify if the AI output is appropriate.} 
The future of CS education needs to equip students with the ``soft skills'' or competencies to work together with AI. \textcolor{pink}{The faculty workshop proposed the Breadcrumb strategy, which embeds these skills throughout the four-year curriculum. Breadcrumbs can be operationalized as small prompts and practical exercises in lectures, assignments, exams, or projects, reinforcing students' AI-native skills year by year in different courses.} Under this theme, we discuss skills for working with AI professionally and ethically, as well as for collaborating with AI efficiently.



\subsubsection{Working Professionally and Ethically}
\textbf{Ethics and societal responsibility as core CS competencies.}
Workshop participants emphasized that ethics, societal impact, and professional responsibility must be treated as foundational skills rather than peripheral considerations. Teaching approaches could, therefore, ensure that students routinely consider the needs of users (including direct users and those otherwise affected or impacted by the systems), accessibility, legal and regulatory constraints, environmental impact, and broader societal consequences in everyday system design and implementation. Ethical and risk-related concerns could be introduced early and revisited throughout the curriculum, reinforcing that responsible judgment about trade-offs, risks, and accountability is inseparable from technical competence, and that both practical and collaborative approachesare often central to addressing such issues.

\textbf{Communication, specification, and justification.}
\textcolor{blue}{Clear communication was identified as central to AI-era computing. As AI increasingly handles implementation, the human bottleneck shifts from writing code to articulating intent: deciding \textit{what} to build, \textit{why} it should be built, \textit{for whom}, and whether the result is correct. This change transforms communication from a secondary skill into a primary method of engineering control. In practical terms, this involves clearly defining problems, intentions, and constraints. This can be done through natural language prompts for AI tools, formal requirements for human collaborators, or explanations for stakeholders.} Solutions involve developing students’ ability to specify problems precisely (for example, by identifying and resolving potential ambiguities in problem statements \cite{denny2025probing}), articulating system requirements, justifing design decisions, and explaining systems and outcomes in accessible language. These skills support collaboration with both humans and AI systems and are essential for ensuring that the ``right problem'' is being solved, not merely that a system functions. 

\textbf{Teamwork and professional practice.}
Participants stressed that collaboration, teamwork, and stakeholder engagement are core CS practices. Teaching solutions could make these skills explicit and assessable\textcolor{blue}{, rather than assuming they emerge implicitly through group projects. Professional skills shall be intentionally cultivated.}

\textcolor{blue}{\textbf{Interdisciplinary engagement and diverse perspectives.}
As AI systems increasingly affect domains beyond computing, CS graduates must be prepared to collaborate with experts from other fields. This requires not only communication skills but also intellectual humility: the ability to recognize the limits of one’s own disciplinary perspective and to engage meaningfully with non-CS stakeholders who bring essential domain knowledge, ethical frameworks, and lived experience. Teaching solutions can rather create structured opportunities for such cross-disciplinary collaboration, particularly in project-based and capstone settings.}

\textbf{Resilience, responsibility, and systems thinking.}
Graduates should be prepared to reason about failure modes, energy use, and externalities of AI systems.
In safety-critical contexts, this responsibility extends to determining the appropriate level of assurance: whether informal testing suffices or formal guarantees are required, as well as selecting and applying verification methods accordingly.
Solutions should help students understand how systems behave under stress or uncertainty and encourage a sense of responsibility for long-term consequences, not just immediate performance.



\subsubsection{Collaborating with AI}
\textbf{AI as collaborator, not replacement.}
Teaching solutions must prepare students to work productively with AI tools in realistic workflows, including code generation, review, debugging, and design assistance. The focus could shift away from merely using AI to produce results to forming a collaborative relationship with AI tools. This means students not only express their intent and control AI outputs but also are aware of, and make the best use of, AI's capabilities and limitations~\cite{Somanath2023ExploringTC}.

\textbf{Specification and verification of AI outputs.}
A recurring criterion was the importance of students being able to specify desired behavior clearly and to verify and evaluate AI-generated artifacts. This includes understanding when AI outputs are incorrect, incomplete, or misaligned with requirements, and being able to correct or contextualize them. Specification languages, type systems, and contract-based design provide structured vocabularies for precisely expressing desired behaviour, making the gap between intent and AI output explicit and inspectable rather than implicit and ad hoc.
As AI increasingly handles low-level syntax, students must focus on abstraction, problem decomposition, and precise reasoning. Teaching solutions could emphasize clarity of thought across representations—natural language, code, and formal models—while maintaining sufficient awareness of lower-level details to reason about failures when abstractions break down.
Crucially, students must also consider the \textit{purpose} of verification and evaluation in \textit{context}, including the potential risks and consequences associated with the AI outputs. \textcolor{blue}{The convergence of AI code generation and formal verification tools further reinforces this emphasis. As AI-generated code proliferates, formal verification offers a scalable alternative to manual code review, and LLMs are themselves beginning to assist in proof generation~\cite{kleppmann2025fv}. Teaching students to work at this intersection positions them for an emerging AI-assisted verification workflow that may define the next generation of software quality assurance.}

\textbf{Critical thinking and reflection.}
Participants highlighted the need to strengthen students’ ability to question AI outputs, reflect on their own AI use, and maintain attention in AI-rich environments. Reflection on process—not just outcomes—was seen as essential to meaningful learning.



\subsection{Re-centering Fundamental Education}
Participants expressed that the priority for undergraduate education in Years 1-2 is to equip students with durable intellectual skills that will remain relevant despite rapid technological change. Students must develop strong foundations in algorithms, data structures, and systems thinking; learning emphasis should shift from tool-specific knowledge toward abstraction, reasoning, and underlying concepts and principles. Equally important is cultivating computational thinking and the capacity to “learn how to learn,” enabling students to decompose unfamiliar problems, reason algorithmically, and adapt to new domains and technologies over time. Ethical and human-centered thinking could be integrated from the outset, helping students reflect on societal impact, user needs, and responsibility before uncritical automation habits take hold. Finally, early curricula could avoid premature specialization—particularly in fast-moving areas such as machine learning—and instead foster broad conceptual understanding of AI and computational systems, ensuring that students are prepared to make informed choices about specialization in later years rather than being narrowly trained too early. 

\textbf{Emphasizing abstraction, reasoning, and principles.}
The discussion argued early undergraduate education should retain core CS foundations---algorithms, data structures, systems thinking---but shift emphasis toward abstraction, reasoning, and principles that remain stable as tools evolve. Solutions could avoid superficial coverage of many new topics in favor of depth in broadly applicable foundations. 
In particular, computational thinking---abstraction, decomposition, and algorithmic reasoning---should serve as a foundation for adaptability. Educators can explicitly support “learning how to learn,” preparing junior students for problems and domains not explicitly taught.

\textbf{Top-down reverse engineering approaches.} 
Existing fundamental education generally starts with small-scale programming tasks, but real-life, we could integrate larger-scale projects into fundamental education, and analysed via ``reverse-engineering'' methods in CS101 and equivalents. This method would offer assignments that require students to extract specifications and requirements from a completed compiled solution. By working the pedagogy top-down, this will provide opportunities to teach principles and core concepts based on existing examples and customized to the expectations of incoming new generations, who are already relying on AI for problem-solving.

\textbf{Integrating ethics, legal frameworks, and human-centered thinking.}
Faculty could introduce human-centered perspectives, including ethics, societal impact, and legal frameworks, in Year 1 rather than postponing them. Offline or discussion-based learning formats were seen as particularly valuable for cultivating reflection, attention, and ethical reasoning, before heavy reliance on AI tools.

\textbf{Providing breadth whilst avoiding premature specialization.}
Workshop participants cautioned against forcing all students into deep machine learning early. Instead, Year 1–2 education shall ensure a broad understanding of AI principles and system behavior, including its wide applicability across fields such as social science and law, without requiring mastery of rapidly changing techniques. This may include incorporating modules about the core foundations of language models, statistics, and optimization techniques, with examples of applications in a variety of areas. This should allow students to choose additional specializations based on their interests and desired future career paths.

\begin{figure}[!t]
    \centering
    \includegraphics[width=\linewidth]{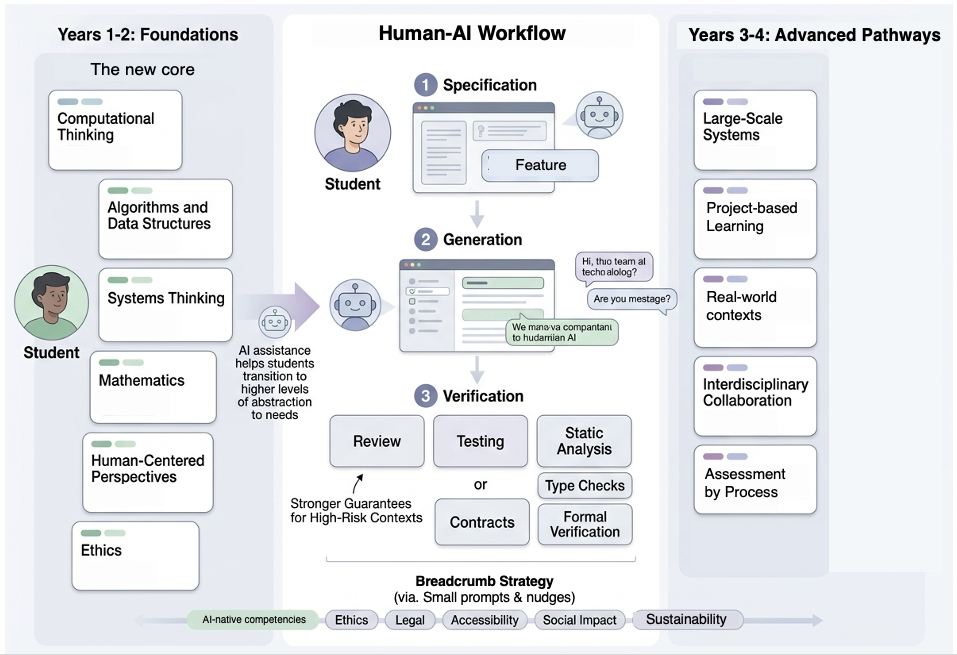}
    \caption{Undergraduate CS curriculum changes must involve recentering fundamental education (Years 1-2) by emphasising principles \& human-centered thinking, and enhancing advanced pathways (Years 3-4) through large-scale, practical, and interdisciplinary project and process-based assessments. Meanwhile, AI-native skills can be integrated seamlessly as Breadcrumbs.}
    \label{fig:curriculum}
\end{figure}

\subsection{Enhancing Advanced Pathways}
In the later years of undergraduate computer science education, workshop participants believed the emphasis should shift from mastering individual techniques to exercising judgment in complex, real-world contexts. Participants highlighted the importance of system-level thinking, with students gaining hands-on experience working with large-scale, distributed, and open-source systems that mirror contemporary software and AI deployments. Project-based and experiential learning—through capstone projects, internships, and applied coursework—was seen as essential for integrating technical depth with ethical reasoning and professional practice. At this stage, students should also be encouraged to engage across disciplinary boundaries, applying computational and AI methods in collaboration with domain experts to address problems beyond traditional computing contexts. Correspondingly, assessment practices must evolve to value process as much as product, evaluating how students reason through design trade-offs, iterate on solutions, reflect on their use of AI tools, and justify their decisions, rather than focusing solely on the final artifact.

\textbf{System-level and large-scale thinking.}
In teaching late undergraduate years, faculty members can emphasize working with large-scale, distributed, and open-source systems that reflect real-world software and AI deployment. Solutions should help students understand how components interact across the computing stack and how AI systems behave in production contexts, including the increasingly service-based nature of modern computing systems and the additional complexities that arise from that.
This can include experience with the quality-assurance challenges specific to AI-generated code, such as detecting subtle increases in code complexity and technical debt, which empirical studies have shown accompany AI tool adoption \cite{he2026cursor}.

\textcolor{blue}{\textbf{Project-based and experiential learning.}}
Criteria strongly supported project-based learning, internships, and applied experiences in Years 3–4. These formats allow students to integrate technical, ethical, and professional skills while working on complex, realistic problems.

\textbf{Interdisciplinary engagement.}
Advanced students should be encouraged to apply AI and computational thinking across domains, collaborating with experts from other fields. This supports transferable skills and reflects how AI is used in practice. Furthermore, computing departments could be positioned as interdisciplinary hubs, allowing cross-domain collaboration. This effort has been piloted in the forms of ``AI+X'' or ``Computing+X'' in various institutions (e.g., NUS\footnote{\url{https://ai.nus.edu.sg/research/ai-x/}}).

\textbf{Assessment of process and judgment.}
Teaching solutions should assess not only final outputs but also students’ processes, including iteration, prompting strategies, reflection on AI use, and decision-making. This helps distinguish genuine learning from superficial AI-assisted production.


\subsection{Embracing New Pedagogies and Teaching Strategies}
Taken together, the shifts toward AI-native competencies and the reform of undergraduate education in both the early and later years make clear that incremental content updates alone are insufficient. If the goals of modern computer science education are to cultivate judgment, adaptability, and responsible human–AI collaboration, then the strategies by which we teach, engage, and assess students must evolve in parallel. The workshop discussions, therefore, converged on the need for a fundamental rethinking of pedagogical approaches—one that aligns institutional realities with these new educational objectives. The following teaching strategies articulate how such changes can be enacted in practice, focusing on scalable, sustainable, and instructor-adoptable approaches that support meaningful learning in the era of AI.

\textbf{Scaling across faculty and institutional constraints.} Effective teaching solutions may be revised scale across courses, institutions, and class sizes. 
Specifically, solutions must be implementable without a complete curriculum redesign. They should be modular, adoptable by individual instructors, and feasible within resource, bureaucratic, and staffing constraints. A core barrier to pedagogical reform, however, is the overwhelming time constraint on faculty, with the current resource model unable to support labor-intensive, process-based assessments. To address this, institutions must make a strategic investment in human resources, beginning with a proactive stance to advocate for increased funding and dedicated personnel. This is essential to sustain diverse and personalized assessment methods. Crucially, administrations must ensure the maintenance of an appropriate instructor-to-student ratio, particularly in foundational courses, to guarantee quality instruction and support the increased grading load associated with new assessment types. 


\textbf{Supporting student motivation, engagement, and autonomy.}
Pedagogy depends on student presence and attention~\cite{Moores02102019}. There is a need to provide clear value for in-person participation and communicate this to students. This can be done by including engaging or enjoyable elements and fostering social and community aspects of learning to mitigate isolation and disengagement. Teaching strategies could allow multiple pathways for students to engage and demonstrate learning, supporting autonomy while maintaining clear standards and assessable outcomes. \textcolor{purple}{Furthermore, persistent and exclusive exposure to AI-generated outputs can quietly erode students' intrinsic motivation, their willingness to experiment, and their confidence in their own intellectual agency. To reach autonomy, students should feel comfortable to take intellectual risks, and understand that while they can offload tasks to AI, learning needs individual’s effort~\cite{charisi2026ai}. Pedagogical designs can therefore create structured spaces where students have the potential to engage independently before turning to AI}

\textbf{Aligning assessment with AI use.}
New pedagogies could encourage responsible and constructive use of AI rather than banning it, and target emerging skills (for example, by having students craft natural language prompts for solving computational tasks, and providing explicit feedback on their quality \cite{padurean2025prompt}). 
\textcolor{blue}{The Prompt Programming platform~\cite{padurean2025prompt} demonstrates one concrete implementation of specification-driven assessment: students must craft natural language specifications to generate code solving specified problems, requiring iterative refinement and critical evaluation of AI outputs. This approach directly instantiates the specification-and-verification workflow advocated in this paper and demonstrates scalability through its freely available implementation.}
Assessment strategies can evaluate learning processes, critical reflection, and effective AI use, ensuring that AI supports rather than undermines educational goals. \textcolor{blue}{Experimental evidence further demonstrates that guardrailed AI deployments, where AI provides structured hints rather than direct solutions, can maintain learning outcomes while preserving the benefits of AI assistance~\cite{bastani2025guardrails}, supporting the principle that AI integration should be carefully designed rather than simply permitted or banned.}
In STEM education, learning can be operationalized as a deep engagement with artifacts and a development of intentionality (i.e., paths to approach an idea)~\cite{Petrich2013ItLL}. Monitoring learning processes would be an effective way to assess whether students are learning. Tool-assisted assessment, where students submit specifications, tests, or formal properties alongside their code, can provide scalable, objective evidence of understanding while naturally encouraging verification-oriented workflows.



\subsection{Shifting Institutional Support}

The successful adoption of AI tools and contemporary teaching methods requires significant institutional support and a strategic reallocation of resources. The faculty workshop identified three critical areas for administrative intervention, focusing on resourcing, incentives, and the creation of shared tools to facilitate scalable and sustained change.


\textbf{Supporting faculty workload.} Institutions can support the use of technology explicitly to reduce mechanical faculty workloads in grading and office-hour duties, redirecting faculty time toward higher-value activities such as curriculum development and personalized student interaction. Furthermore, the faculty or school could establish specialized support services, such as ethics consulting services, to assist both students and faculty in navigating the complex ethical challenges inherent in AI system design.

\textbf{Considerations for Academic Concessions} Traditionally, in many universities, students receive concessions (e.g., extra time to complete an assessment) for compassionate or unforeseen circumstances. Concessions are given under the assumption that students work on their own. However, access to AI tools may violate this assumption. Staff who determine academic concession solutions now need to understand how AI affects learning and exam participation.

\textbf{Incentivizing faculty participation.} For innovative teaching to become widespread, institutions must move beyond passive encouragement to actively incentivize faculty participation. To foster the development and sharing of new materials, a system of financial and professional incentives is recommended, including small awards or grants for contributing high-quality, "AI-ready" assignments and reports to shared repositories. This should be coupled with initiatives to enhance the prestige and visibility of such contributions through internal recognition or publication pathways. 


\textbf{Maximizing efficiency and scalability.} Institutions may invest in establishing a centralized, moderated, open resource repository to host actionable assignments, projects, and reports that accelerate AI integration. A dedicated, small moderation team is necessary to oversee submissions and ensure quality control. To complement this repository, institutions would need to support the organization of an annual assignment-sharing conference to facilitate the global discussion, submission, and adoption of tested, "AI-ready assignments," thus creating a scalable learning ecosystem. This ensures that faculty have immediate access to a "smorgasbord" of practical, embeddable materials, preventing the need for individual course redesigns and fostering a culture of rapid, informed pedagogical adaptation.


\newpage

\section{Conclusion and Next Steps}

The rise of generative AI represents a significant and likely lasting disruption to CS education and the broader IT industry, though the ultimate scope and trajectory of this transformation remain unknown. As AI tools increasingly automate tasks traditionally performed by junior software engineers, the traditional assumptions in CS education -- where coding, debugging, and documentation are scarce -- no longer hold. 
This shift calls an urgent and comprehensive reform of CS curricula, moving away from a focus on manual code generation toward higher-order skills such as system architecting, verification, and ethical responsibility. Early empirical evidence supports this urgency~\cite{he2026cursor}, confirming that the quality assurance gap is not hypothetical but present and measurable. To remain relevant, CS education must adapt with expediency to produce graduates who can operate immediately and steward complex, AI-generated systems.

\textcolor{blue}{In light of recent significant changes, the NUS-Google Workshops brought together a diverse group of faculty and students to gather insights from key stakeholders in CS education. These two workshops—one focusing on faculty and institutional challenges, and the other on student experiences and concerns—offer several valuable contributions that enhance existing initiatives. 
Specifically, the dual-workshop approach captures both faculty and student perspectives, highlighting areas of agreement and disagreement that single-stakeholder methods overlook. In addition, rather than focusing solely on course-level responses, this paper proposes criteria for restructuring programs across all four years of undergraduate study.} Both groups affirmed the enduring value of the foundation knowledge while stressing the paramount importance of metacognitive skills such as computational thinking, critical verification, accountability, and ethical reasoning. We note that ``verification'' as a graduate competency spans a wide range of rigour—from code review and testing to type-system-guided development and formal proof—and that curriculum designers can be more explicit about which levels of this spectrum they aim to teach and assess.

This white paper outlines a strategic response and documents several concrete solutions proposed during the workshops, such as "prompt-first" or "reverse-engineering" courses, new project-based learning models, and the ``breadcrumb`` strategy to embed and reinforce AI-native skills throughout the curriculum. However, the efficacy of these proposed reforms remains an open question. Future work must be rigorously dedicated to validating these solutions against the key criteria and requirements voiced by the workshop attendees. Furthermore, to fully align the curriculum with the rapidly evolving job market, future studies will require the critical input of a broader range of industry practitioners and business stakeholders to ensure that academic pathways genuinely prepare the next generation of engineers for long-term success. In particular, the Practical Solutions envisioned by the proposed criteria (Section~\ref{criteria}) remain to be implemented; we invite the community to contribute concrete, implemented examples—such as specification-driven assignments, verification-oriented project workflows, and AI-integrated capstone designs—against which the criteria in this paper can be tested. The evolving capabilities of AI and how they are integrated in the industry also mean this will require constant re-evaluation of the CS curricula over time.




\newpage

\section{Whitepaper Contributors}

\begin{itemize}
    \item Yi-Chieh (EJ) Lee, National University of Singapore, yclee@nus.edu.sg \footnote{The author ordering reflects the chronological sequence in which authors entered and edited the manuscript, rather than relative contribution.}
    \item Nattapat Boonprakong, National University of Singapore, nattapat.boonprakong@nus.edu.sg
    \item Yugin Tan, National University of Singapore, tanyugin@nus.edu.sg
    \item Harold Soh, National University of Singapore, dcshssh@nus.edu.sg
    \item Alex Potanin, Australian National University, alex.potanin@anu.edu.au
    \item Viraj Kumar, Indian Institute of Science, viraj@iisc.ac.in
    \item Anoop K. Sinha, Google, Paradigms of Intelligence, anoopsinha@google.com
    \item Chen Qian, Shanghai Jiao Tong University, qianc@sjtu.edu.cn
    \item Paul Denny, University of Auckland, paul@cs.auckland.ac.nz 
    \item Mennatallah El-Assady, ETH Zurich, melassady@ai.ethz.ch
    \item Ian Oakley, KAIST, ian.r.oakley@gmail.com
    \item Jake Renzella, University of New South Wales, jake.renzella@unsw.edu.au
    \item Amy Zhang, University of Washington, axz@cs.uw.edu
    \item Jat Singh, RC-Trust, UA-Ruhr\slash UDE \& University of Cambridge, jatinder.singh@cst.cam.ac.uk
    \item Wee Sun Lee, National University of Singapore, dcsleews@nus.edu.sg
    \item Hsuan-Tien Lin, National Taiwan University, htlin@csie.ntu.edu.tw
    \item Jane L. E, National University of Singapore, ejane@nus.edu.sg
    \item Anthony Tang, Singapore Management University, tonyt@smu.edu.sg
    \item Margaret M. Burnett, Oregon State University, burnett@eecs.oregonstate.edu
    \item Sowmya Somanath, University of Victoria, sowmyasomanath@uvic.ca 
    \item Renwen Zhang, Nanyang Technological University, renwen.zhang@ntu.edu.sg
    \item Vicky Charisi, Singapore-MIT Alliance for Research and Technology, vasiliki.charisi@smart.mit.edu
    \item Alexandra I. Cristea, University of Durham, alexandra.i.cristea@durham.ac.uk


\end{itemize}

\section{Workshop Organizers}
\subsection{Faculty Workshop Organizers}
\begin{itemize}
    \item Yi-Chieh (EJ) Lee, National University of Singapore
    \item Zicheng Zhu, National University of Singapore
    \item Michael Terry, Google
    \item Shruti Sheth, Google
    \item Anoop K. Sinha, Google

    \item Harold Soh, National University of Singapore
\end{itemize}

\subsection{Student Workshop Organizers}

\begin{itemize}
    \item Nattapat Boonprakong, National University of Singapore
    \item Jingshu Li, National University of Singapore
    \item Yugin Tan, National University of Singapore
    \item Vishruti Ranjan, National University of Singapore
    \item Emran Poh, Singapore Management University
    \item Tianqi Song, National University of Singapore
    \item Peinuan Qin, National University of Singapore
    \item Yitian Yang, National University of Singapore
    \item Zhengtao Xu, National University of Singapore
    \item Han Meng, National University of Singapore
    \item Junti Zhang, National University of Singapore
    \item Xuehui Yu, National University of Singapore
    \item Pranav Dangi, National University of Singapore
    \item Weiyan Shi, Singapore University of Technology and Design

\end{itemize}

\textbf{Our workshops were made possible through the generous support of Google.}

\section{Faculty Workshop Attendees}
\begin{itemize}
\item Alexandra I. Cristea, Durham University
\item Alex Potanin, Australian National University
\item Amy X. Zhang, University of Washington
\item Anoop K. Sinha, Google
\item April Yi Wang, ETH Zürich
\item Chen Qian, Shanghai Jiao Tong University
\item Chris Piech, Stanford University
\item Jane L. E, National University of Singapore
\item Hsuan-Tien Lin, National Taiwan University
\item Ian Oakley, KAIST
\item Jake Renzella, University of New South Wales
\item Li-Shiuan Peh, National University of Singapore
\item Jat Singh, RC-Trust, UA-Ruhr\slash UDE \& University of Cambridge
\item Johannes Schöning, University of St. Gallen
\item Khai N. Truong, University of Toronto
\item Paul Denny, The University of Auckland
\item Margaret M. Burnett, Oregon State University
\item Mennatallah El-Assady, ETH Zürich
\item Wei Tsang Ooi, National University of Singapore
\item Sowmya Somanath, University of Victoria
\item Xiaojuan Ma, Hong Kong University of Science and Technology
\item Robert Xiao, University of British Columbia
\item Roman Cunci, Avature
\item Weinan E, Peking University
\item Ye Wang, National University of Singapore
\item Tony Tang, Singapore Management University
\item Viraj Kumar, IISc Bangalore
\item Wee Sun Lee, National University of Singapore
\item Yi-Chieh Lee, National University of Singapore
\end{itemize}

\section{Student Workshop Attendees}
\begin{itemize}
\item Alex Chien, 	National University of Singapore
\item Arpita Dutta, 	National University of Singapore
\item Aryan Jain, 	National University of Singapore
\item Botao 'Amber' Hu, 	University of Oxford
\item Chan Lok Tung, 	National University of Singapore
\item Chayapol Chaoveeraprasit, 	National University of Singapore
\item Emily Aurelia, 	Singapore Management University
\item G Dhanesh Sabaratnam, 	Singapore Institute of Management-London School of Economics
\item Gangyi Zhang, 	The University of Melbourne
\item Grant Lee, 	ST Engineering
\item Gu Zhen Feng, 	National University of Singapore
\item Gullapalli Chetan, 	National University of Singapore
\item Hans, 	Singapore Polytechnic
\item Ho Yu Hua, 	Singapore University of Technology and Design
\item Irene Ho, 	University of Wisconsin-Madison
\item Jason Pang, 	University of Pennsylvania
\item Jessica Chen, 	National University of Singapore
\item Jingzhu Chen, 	Tongji University
\item King Jean Lynn, 	Nanyang Technological University
\item Lim Jie Han, 	Nanyang Technological University
\item Liu Yan, 	National University of Singapore
\item Meng Yuxuan, 	National University of Singapore
\item Michael Cenreng, 	National University of Singapore
\item Ong Jia Cheng, 	National University of Singapore
\item Pang Zi Haur, 	Kyoto University
\item Pannapat Chanpaisaeng, 	Singapore University of Technology and Design
\item Phua Yu Cheng, 	Universiti Teknologi Malaysia
\item Rachel Yeo Hui Min, 	National University of Singapore
\item Rimmon Saloman Bhosale, 	Singapore Management University
\item Sarthak Kumar, 	National University of Singapore
\item Shizun Wang, 	National University of Singapore
\item Sparsh Rastogi, 	Thapar Institute of Engineering and Technology 
\item Tan Jin Hong, 	Universiti Tunku Abdul Rahman
\item Tee Qian Hui, 	Universiti Tunku Abdul Rahman
\item Ting Zhang, 	National University of Singapore
\item Vignesh Sundararaj, 	Singapore Management University
\item Wang Chuansheng, 	National University of Singapore
\item Wang Qilin, 	University of Wollongong
\item Weiyan Shi, 	Singapore University of Technology and Design
\item Yang Chen Lin, 	National Tsing Hua University
\item Yew Foong Yik, 	Asia Pacific University of Technology \& Innovation
\item Yifei Luo, 	RWTH Aachen
\item Yiran Jiang, 	Dartmouth College
\item Zhang Dingpu, 	Nanyang Technological University
\item Zhang Jingze, 	National University of Singapore
\item Zhicong Chen, 	National University of Singapore
\item Zhizhou Cao, 	National University of Singapore
\end{itemize}

\bibliography{references.bib}

@inproceedings{lee2025impact,
  title={The impact of generative AI on critical thinking: Self-reported reductions in cognitive effort and confidence effects from a survey of knowledge workers},
  author={Lee, Hao-Ping and Sarkar, Advait and Tankelevitch, Lev and Drosos, Ian and Rintel, Sean and Banks, Richard and Wilson, Nicholas},
  booktitle={Proceedings of the 2025 CHI Conference on Human Factors in Computing Systems},
  pages={1--22},
  year={2025}
}

@inproceedings{tankelevitch2024metacognitive,
  title={The metacognitive demands and opportunities of generative {AI}},
  author={Tankelevitch, Lev and Kewenig, Viktor and Simkute, Auste and Scott, Ava Elizabeth and Sarkar, Advait and Sellen, Abigail and Rintel, Sean},
  booktitle={CHI Conference on Human Factors in Computing Systems},
  pages={1--24},
  year={2024}
}

@inproceedings{choudhuri2025genAIstudents,
  title={{Insights from the Frontline: GenAI Utilization Among Software Engineering Students}},
  author={Choudhuri, Rudrajit and Ramakrishnan, Ambareesh and Chatterjee, Amreeta and Trinkenreich, Bianca and Steinmacher, Igor and Gerosa, Marco and Sarma, Anita},
  booktitle={2025 IEEE/ACM 37th International Conference on Software Engineering Education and Training (CSEE\&T)},
  pages={1--12},
  year={2025},
  organization={IEEE}
}

@article{fan2025metacoglazy,
  title={Beware of metacognitive laziness: Effects of generative artificial intelligence on learning motivation, processes, and performance},
  author={Fan, Yizhou and Tang, Luzhen and Le, Huixiao and Shen, Kejie and Tan, Shufang and Zhao, Yueying and Shen, Yuan and Li, Xinyu and Ga{\v{s}}evi{\'c}, Dragan},
  journal={British Journal of Educational Technology},
  volume={56},
  number={2},
  pages={489--530},
  year={2025},
  publisher={Wiley Online Library}
}

@article{choudhuri2026cantthink,
  title={Why Johnny Can't Think: GenAI's Impacts on Cognitive Engagement},
  author={Choudhuri, Rudrajit and Sanchez, Christopher and Burnett, Margaret and Sarma, Anita},
  journal={arXiv preprint arXiv:2601.22430},
  year={2026}
}

@misc{computer2026perplexity,
title={Introducing Perplexity Computer},
url={https://www.perplexity.ai/hub/blog/introducing-perplexity-computer},
author={Perplexity AI}
}

@misc{phd2025bbc,
title={OpenAI claims GPT-5 model boosts ChatGPT to 'PhD level'},
url={https://www.bbc.com/news/articles/cy5prvgw0r1o},
author={BBC}
}

@inproceedings{10.1145/3756681.3757081,
author = {Yu, Liang},
title = {Paradigm shift on Coding Productivity Using GenAI},
year = {2025},
isbn = {9798400713859},
publisher = {Association for Computing Machinery},
address = {New York, NY, USA},
url = {https://doi.org/10.1145/3756681.3757081},
doi = {10.1145/3756681.3757081},
booktitle = {Proceedings of the 29th International Conference on Evaluation and Assessment in Software Engineering},
pages = {708–713},
numpages = {6},
location = {
},
series = {EASE '25}
}

@misc{peng2023impactaideveloperproductivity,
      title={The Impact of AI on Developer Productivity: Evidence from GitHub Copilot}, 
      author={Sida Peng and Eirini Kalliamvakou and Peter Cihon and Mert Demirer},
      year={2023},
      eprint={2302.06590},
      archivePrefix={arXiv},
      primaryClass={cs.SE},
      url={https://arxiv.org/abs/2302.06590}, 
}

@article{denny2024computing,
author = {Denny, Paul and Prather, James and Becker, Brett A. and Finnie-Ansley, James and Hellas, Arto and Leinonen, Juho and Luxton-Reilly, Andrew and Reeves, Brent N. and Santos, Eddie Antonio and Sarsa, Sami},
title = {Computing Education in the Era of Generative AI},
year = {2024},
issue_date = {February 2024},
publisher = {Association for Computing Machinery},
address = {New York, NY, USA},
volume = {67},
number = {2},
issn = {0001-0782},
url = {https://doi.org/10.1145/3624720},
doi = {10.1145/3624720},
journal = {Commun. ACM},
month = jan,
pages = {56–67},
numpages = {12}
}

@inproceedings{clear2025ai,
author = {Clear, Tony and Cajander, \AAsa and Clear, Alison and McDermott, Roger and Daniels, Mats and Divitini, Monica and Forshaw, Matthew and Humble, Niklas and Kasinidou, Maria and Kleanthous, Styliani and Kultur, Can and Parvini, Ghazaleh and Polash, Mohammad and Zhu, Tingting},
title = {AI Integration in the IT Professional Workplace: A Scoping Review and Interview Study with Implications for Education and Professional Competencies},
year = {2025},
isbn = {9798400712081},
publisher = {Association for Computing Machinery},
address = {New York, NY, USA},
url = {https://doi.org/10.1145/3689187.3709607},
doi = {10.1145/3689187.3709607},
booktitle = {2024 Working Group Reports on Innovation and Technology in Computer Science Education},
pages = {34–67},
numpages = {34},
location = {Milan, Italy},
series = {ITiCSE 2024}
}

@inproceedings{prather2025beyond,
author = {Prather, James and Leinonen, Juho and Kiesler, Natalie and Gorson Benario, Jamie and Lau, Sam and MacNeil, Stephen and Norouzi, Narges and Opel, Simone and Pettit, Vee and Porter, Leo and Reeves, Brent N. and Savelka, Jaromir and Smith, David H., IV and Strickroth, Sven and Zingaro, Daniel},
title = {Beyond the Hype: A Comprehensive Review of Current Trends in Generative AI Research, Teaching Practices, and Tools},
year = {2025},
isbn = {9798400712081},
publisher = {Association for Computing Machinery},
address = {New York, NY, USA},
url = {https://doi.org/10.1145/3689187.3709614},
doi = {10.1145/3689187.3709614},
booktitle = {2024 Working Group Reports on Innovation and Technology in Computer Science Education},
pages = {300–338},
numpages = {39},
location = {Milan, Italy},
series = {ITiCSE 2024}
}

@inproceedings{dijkstra1978foolishness,
author = {Dijkstra, Edsger W.},
title = {On the Foolishness of "Natural Language Programming"},
year = {1978},
isbn = {354009251X},
publisher = {Springer-Verlag},
address = {Berlin, Heidelberg},
booktitle = {Program Construction, International Summer Schoo},
pages = {51–53},
numpages = {3}
}

@inproceedings{gries1974what,
author = {Gries, David},
title = {What should we teach in an introductory programming course?},
year = {1974},
isbn = {9781450374835},
publisher = {Association for Computing Machinery},
address = {New York, NY, USA},
url = {https://doi.org/10.1145/800183.810447},
doi = {10.1145/800183.810447},
booktitle = {Proceedings of the Fourth SIGCSE Technical Symposium on Computer Science Education},
pages = {81–89},
numpages = {9},
series = {SIGCSE '74}
}

@book{swebok2024,
  editor    = {Hironori Washizaki},
  title     = {Guide to the Software Engineering Body of Knowledge (SWEBOK Guide)},
  edition   = {4.0},
  year      = {2024},
  publisher = {IEEE Computer Society},
  url       = {https://www.swebok.org}
}

@inproceedings{prather2023robots,
author = {Prather, James and Denny, Paul and Leinonen, Juho and Becker, Brett A. and Albluwi, Ibrahim and Craig, Michelle and Keuning, Hieke and Kiesler, Natalie and Kohn, Tobias and Luxton-Reilly, Andrew and MacNeil, Stephen and Petersen, Andrew and Pettit, Raymond and Reeves, Brent N. and Savelka, Jaromir},
title = {The Robots Are Here: Navigating the Generative AI Revolution in Computing Education},
year = {2023},
isbn = {9798400704055},
publisher = {Association for Computing Machinery},
address = {New York, NY, USA},
url = {https://doi.org/10.1145/3623762.3633499},
doi = {10.1145/3623762.3633499},
booktitle = {Proceedings of the 2023 Working Group Reports on Innovation and Technology in Computer Science Education},
pages = {108–159},
numpages = {52},
location = {Turku, Finland},
series = {ITiCSE-WGR '23}
}

@inproceedings{shihab2025effects,
author = {Shihab, Md Istiak Hossain and Hundhausen, Christopher and Tariq, Ahsun and Haque, Summit and Qiao, Yunhan and Mulanda, Brian Wise},
title = {The Effects of GitHub Copilot on Computing Students' Programming Effectiveness, Efficiency, and Processes in Brownfield Coding Tasks},
year = {2025},
isbn = {9798400713408},
publisher = {Association for Computing Machinery},
address = {New York, NY, USA},
url = {https://doi.org/10.1145/3702652.3744219},
doi = {10.1145/3702652.3744219},
booktitle = {Proceedings of the 2025 ACM Conference on International Computing Education Research V.1},
pages = {407–420},
numpages = {14},
location = {
},
series = {ICER '25}
}

@inproceedings{denny2025probing,
author = {Denny, Paul and Kumar, Viraj and MacNeil, Stephen and Prather, James and Leinonen, Juho},
title = {Probing the Unknown: Exploring Student Interactions with Probeable Problems at Scale in Introductory Programming},
year = {2025},
isbn = {9798400715679},
publisher = {Association for Computing Machinery},
address = {New York, NY, USA},
url = {https://doi.org/10.1145/3724363.3729093},
doi = {10.1145/3724363.3729093},
booktitle = {Proceedings of the 30th ACM Conference on Innovation and Technology in Computer Science Education V. 1},
pages = {618–624},
numpages = {7},
location = {Nijmegen, Netherlands},
series = {ITiCSE 2025}
}

@inproceedings{padurean2025prompt,
author = {P\u{a}durean, Victor-Alexandru and Denny, Paul and Gotovos, Alkis and Singla, Adish},
title = {Prompt Programming: A Platform for Dialogue-based Computational Problem Solving with Generative AI Models},
year = {2025},
isbn = {9798400715679},
publisher = {Association for Computing Machinery},
address = {New York, NY, USA},
url = {https://doi.org/10.1145/3724363.3729094},
doi = {10.1145/3724363.3729094},
booktitle = {Proceedings of the 30th ACM Conference on Innovation and Technology in Computer Science Education V. 1},
pages = {458–464},
numpages = {7},
location = {Nijmegen, Netherlands},
series = {ITiCSE 2025}
}

@inproceedings{he2026cursor,
  author    = {Hao He and Courtney Miller and Shyam Agarwal and Christian K{\"a}stner and Bogdan Vasilescu},
  title     = {Speed at the Cost of Quality: How {Cursor AI} Increases Short-Term Velocity and Long-Term Complexity in Open-Source Projects},
  booktitle = {Proceedings of the 23rd International Conference on Mining Software Repositories},
  series    = {MSR '26},
  year      = {2026},
  location  = {Rio de Janeiro, Brazil},
  publisher = {ACM},
  address   = {New York, NY, USA},
  doi       = {10.1145/3793302.3793349},
  note      = {arXiv:2511.04427}
}

@book{Reddi2026MLSystems,
  author = {Vijay Janapa Reddi},
  title = {{Introduction to Machine Learning Systems}},
  year = {2026},
  publisher = {{MIT Press}},
  note = {Available online at \url{https://mlsysbook.ai/book/}}
}

@article{Moores02102019,
author = {Elisabeth Moores and Gurkiran K. Birdi and Helen E. Higson},
title = {Determinants of university students’ attendance},
journal = {Educational Research},
volume = {61},
number = {4},
pages = {371--387},
year = {2019},
publisher = {Routledge},
doi = {10.1080/00131881.2019.1660587},


URL = { 
    
        https://doi.org/10.1080/00131881.2019.1660587
    
    

},
eprint = { 
    
        https://doi.org/10.1080/00131881.2019.1660587
    
    

}

}

@inproceedings{Petrich2013ItLL,
  title={It Looks Like Fun, but Are They Learning?},
  author={Mike Petrich and Karen Wilkinson and Bronwyn Bevan},
  year={2013},
  url={https://api.semanticscholar.org/CorpusID:111722217}
}

@inproceedings{Somanath2023ExploringTC,
  title={Exploring the Composite Intentionality of 3D Printers and Makers in Digital Fabrication},
  author={Sowmya Somanath and Ron Wakkary and Omid Ettehadi and Henry W. J. Lin and Armi Behzad and Jordan Eshpeter and Doenja Oogjes},
  year={2023},
  url={https://api.semanticscholar.org/CorpusID:261364233}
}

@article{mohammadkarimi2023teachers,
  title={Teachers' reflections on academic dishonesty in EFL students' writings in the era of artificial intelligence},
  author={Mohammadkarimi, Ebrahim},
  journal={Journal of Applied Learning \& Teaching},
  volume={6},
  number={2},
  pages={105--113},
  year={2023},
  publisher={Kaplan Business School Australia Sydney, NSW}
}

@article{Theobold04052021,
author = {Allison S. Theobold},
title = {Oral Exams: A More Meaningful Assessment of Students’ Understanding},
journal = {Journal of Statistics and Data Science Education},
volume = {29},
number = {2},
pages = {156--159},
year = {2021},
publisher = {Taylor \& Francis},
doi = {10.1080/26939169.2021.1914527},


URL = { 
    
        https://doi.org/10.1080/26939169.2021.1914527
    
    

},
eprint = { 
    
        https://doi.org/10.1080/26939169.2021.1914527
    
    

}

}

@article{fmthinking2024,
  title={On formal methods thinking in computer science education},
  author={Dongol, Brijesh and Dubois, Catherine and Hallerstede, Stefan and Hehner, Eric and Morgan, Carroll and M{\"u}ller, Peter and Ribeiro, Leila and Silva, Alexandra and Smith, Graeme and de Vink, Erik},
  journal={Formal Aspects of Computing},
  volume={37},
  number={1},
  pages={1--23},
  year={2024},
  publisher={ACM New York, NY}
}

@inproceedings{eaton2024aics2023,
  title={Artificial intelligence in the cs2023 undergraduate computer science curriculum: Rationale and challenges},
  author={Eaton, Eric and Epstein, Susan L},
  booktitle={Proceedings of the AAAI Conference on Artificial Intelligence},
  volume={38},
  number={21},
  pages={23078--23083},
  year={2024}
}

@misc{cs2023curriculum,
  title={Computer science curricula 2023},
  author={Kumar, Amruth N and Raj, Rajendra K and Aly, Sherif G and Anderson, Monica D and Becker, Brett A and Blumenthal, Richard L and Eaton, Eric and Epstein, Susan L and Goldweber, Michael and Jalote, Pankaj and others},
  year={2024},
  publisher={ACM}
}

@article{becker2025metr,
  title={Measuring the impact of early-2025 AI on experienced open-source developer productivity},
  author={Becker, Joel and Rush, Nate and Barnes, Elizabeth and Rein, David},
  journal={arXiv preprint arXiv:2507.09089},
  year={2025}
}

@article{bastani2025guardrails,
  title={Generative AI without guardrails can harm learning: Evidence from high school mathematics},
  author={Bastani, Hamsa and Bastani, Osbert and Sungu, Alp and Ge, Haosen and Kabakc{\i}, {\"O}zge and Mariman, Rei},
  journal={Proceedings of the National Academy of Sciences},
  volume={122},
  number={26},
  pages={e2422633122},
  year={2025},
  publisher={National Academy of Sciences}
}

@article{brynjolfsson2025canaries,
  title={Canaries in the Coal Mine?: Six Facts about the Recent Employment Effects of Artificial Intelligence},
  author={Brynjolfsson, Erik and Chandar, Bharat and Chen, Ruyu},
  year={2025},
  publisher={Stanford Institute for Economic Policy Research (SIEPR)}
}

@misc{amasanti2025impactaigeneratedsolutionssoftware,
      title={The Impact of AI-Generated Solutions on Software Architecture and Productivity: Results from a Survey Study}, 
      author={Giorgio Amasanti and Jasmin Jahic},
      year={2025},
      eprint={2506.17833},
      archivePrefix={arXiv},
      primaryClass={cs.SE},
      url={https://arxiv.org/abs/2506.17833}, 
}

@misc{ipeirotis2026scalablepersonalizedoralassessments,
      title={Scalable and Personalized Oral Assessments Using Voice AI}, 
      author={Panos Ipeirotis and Konstantinos Rizakos},
      year={2026},
      eprint={2603.18221},
      archivePrefix={arXiv},
      primaryClass={cs.CY},
      url={https://arxiv.org/abs/2603.18221}, 
}

@online{kleppmann2025fv,
  author = {Martin Kleppmann},
  title = {Prediction: AI will make formal verification go mainstream},
  year = 2025,
  url = {https://martin.kleppmann.com/2025/12/08/ai-formal-verification.html},
  urldate = {2025-12-08}
}

@techreport{charisi2026ai,
  author      = {Charisi, Vasiliki},
  title       = {Artificial Intelligence in Classrooms: Pedagogical Dimensions},
  institution = {European Parliament, Policy Department for Citizens, Equality and Culture},
  year        = {2026},
  month       = {March},
  number      = {PE 784.574},
  type        = {Briefing},
  url         = {https://www.europarl.europa.eu/RegData/etudes/BRIE/2026/784574/IUST_BRI(2026)784574_EN.pdf}
}

\end{document}